\documentclass[prb,twocolumn,aps,superscriptaddress]{revtex4-2}

\usepackage{graphicx}
\usepackage{dcolumn}
\usepackage{bm}
\usepackage{amssymb}
\usepackage{amsmath}
\usepackage{mathtools}

\usepackage{physics}
\usepackage{sublabel}
\usepackage[utf8]{inputenc}
\usepackage{hyperref}
\usepackage[usenames, dvipsnames]{color}
\usepackage[english]{babel}
\usepackage[T1]{fontenc}
\usepackage{bbm}

\usepackage{float}
\usepackage{verbatim}
\usepackage[caption=false]{subfig}
\usepackage{xfrac}
\usepackage{upgreek}
\usepackage{makecell}
\usepackage{comment}
\usepackage{ulem}

\hypersetup{colorlinks = true, linkcolor=blue, citecolor=blue, urlcolor=blue}

  \makeatletter
    \renewcommand\@make@capt@title[2]{%
     \@ifx@empty\float@link{\@firstofone}{\expandafter\href\expandafter{\float@link}}%
      {\textsc{#1}}\@caption@fignum@sep#2\quad}%
    \makeatother

\newcommand{\KP}{K^\prime}
\newcommand{\kup}{K\uparrow}
\newcommand{\kdown}{K\downarrow}
\newcommand{\kpup}{K^\prime\uparrow}
\newcommand{\kpdown}{K^\prime\downarrow}

\begin{document}

\title{Unifying recent experiments on spin-valley locking in TMDC quantum dots }

\author{Aakash Shandilya}
\email{denotes equal contribution}
\affiliation{Department of Physics, Indian Institute of Technology Bombay, Powai, Mumbai-400076, India}

\author{Sundeep Kapila}
\email{denotes equal contribution}
\affiliation{Department of Electrical Engineering, Indian Institute of Technology Bombay, Powai, Mumbai-400076, India}

\author{Radha Krishnan}
\email{Current Address: Hybrid Quantum Circuits Laboratory, Institute of Physics, \'Ecole Polytechnique F\'ed\'erale de Lausanne (EPFL), Lausanne 1015, Switzerland.}
\affiliation{Division of Physics and Applied Physics, School of Physical and Mathematical Sciences, Nanyang Technological University, Singapore 637371, Singapore}

\author{Bent Weber}
\email{Corresponding author: b.weber@ntu.edu.sg}
\affiliation{Division of Physics and Applied Physics, School of Physical and Mathematical Sciences, Nanyang Technological University, Singapore 637371, Singapore}

\author{Bhaskaran Muralidharan}
\email{Corresponding author: bm@ee.iitb.ac.in}
\affiliation{Department of Electrical Engineering, Indian Institute of Technology Bombay, Powai, Mumbai-400076, India} 
\affiliation{Centre of Excellence in Quantum Information, Computation, Science and Technology, Indian Institute of Technology Bombay, Powai, Mumbai-400076, India}

\date{\today}

\begin{abstract}
The spin-valley or Kramers qubit promises significantly enhanced spin-valley lifetimes due to strong coupling of the electrons' spin to their momentum (valley) degrees of freedom. In transition metal dichalcogenides (TMDCs) such spin-valley locking is expected to be particularly strong owing to the significant intrinsic spin-orbit coupling strength. Very recently, a small number of experiments on TMDC quantum dots have put forth evidence for spin-valley locking for the first time at the few-electron limit. Employing quantum transport theory, here we numerically simulate their ground- and excited-state transport spectroscopy signatures in a unified theoretical framework. In doing so, we reveal the operating conditions under which spin-valley locking occurs in TMDC quantum dots, thereby weaving the connection between intrinsic material properties and the experimental data under diverse conditions. Our simulations thus provide a predictive modeling tool for TMDC quantum dots at the few-electron limit allowing us to deduce from experiments the degree of spin-valley locking based on the SOC strength, inter-valley mixing, and the spin and valley $g$-factors. Our theoretical analysis provides an important milestone towards the next challenge of experimentally confirming valley-relaxation times using single-shot projective measurements. 
\end{abstract}

\maketitle


\begin{figure*}[thpb]
\centering
\includegraphics[width=\textwidth]{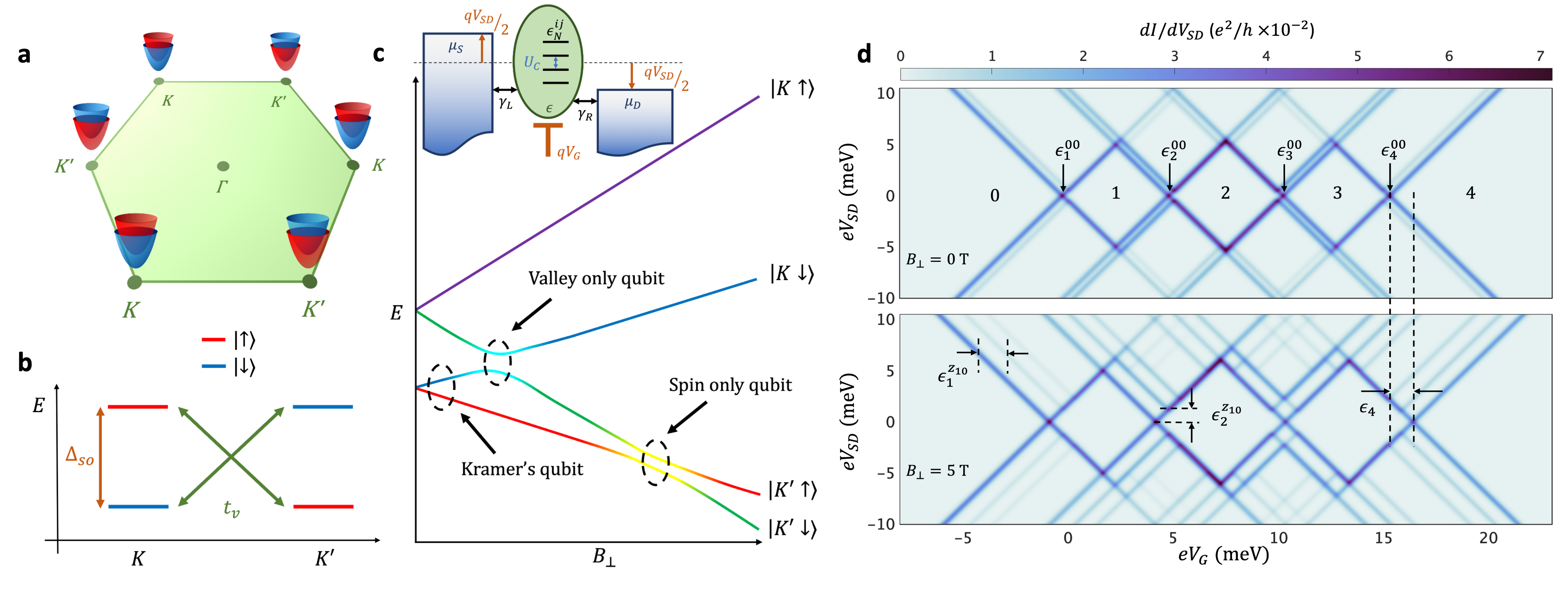}

\captionsetup{justification=justified,singlelinecheck=false}
\caption[width=\textwidth]{
Preliminaries : \textbf{(a)} A schematic illustrating the first Brillouin zone of the TMDC lattice and spin-valley locking at the Dirac points $K$ and $\KP$ at $B=0$. \textbf{(b)} Schematic showing the four conduction band ground states of the system at $B=0$. Here, $\Delta_{so}$ is the spin-orbit splitting energy of each valley and $t_v$ determines the strength of intervalley mixing. \textbf{(c)} Evolution of various one electron energy states in an out-of-plane magnetic field, that can enable different kinds of qubit systems. (Inset) Schematic of the device structure with the transport channels, probed via the transport spectroscopy experiments. The bias potential $V_{SD}$ controls the source-drain bias and $V_G$ is the gate voltage or the plunger voltage that controls the onsite energy. The meaning of each symbol is described in the main text. \textbf{(d)} Coulomb diamonds in the absence and the presence of an out of plane magnetic field. The quantity $\epsilon_N^{00}$ is the transport channel corresponding to the transition from $(N-1)$ electron(s) ground state to $N$ electron(s) ground state in the dot. 
}
\label{fig:fig_1}
\end{figure*}

\section{Introduction\label{sec:introdution}}
Semiconductor quantum dots (QD) are being actively pursued as components in scalable solid-state quantum computing architectures \cite{liu20192d,PhysRevB.101.035204,PhysRevB.104.085421,PhysRevResearch.4.013039}. Realizing QDs in conventional covalent semiconductors such as Si and/or Ge 
\cite{10.1093/nsr/nwy153,RevModPhys.79.1217,RevModPhys.95.025003} promises scalability due to the well-developed fabrication techniques at an industrial scale, while weak hyperfine and spin-orbit interactions \cite{RevModPhys.85.961} permit comparatively long spin lifetimes and coherence times \cite{yoneda2018quantum,PhysRevB.104.085421}. Recently, QDs in two-dimensional (2D) semiconductors such as bilayer graphene (BLG) and the transition-metal dichalcogenides (TMDCs) have generated a lot of interest. In these materials, the presence of non-equivalent valleys can allow for new qubit types and control \cite{kormanyos2014spin,szechenyi2018impurity}, including spin-, valley-, and spin-valley (Kramers) qubits. In particular, the nonvanishing Berry curvature near the degeneracy points \cite{PhysRevB.101.195406,PhysRevLett.125.116804,mccann2013electronic,li2014valley,srivastava2015valley} couples spin to valley
\cite{knothe2018influence,eich2018coupled,banszerus2020electron}, which can, in the presence of spin-orbit coupling (SOC), create a conjugate pair of Kramers doublets $(\{\ket{\kup},\ket{\kpdown}\}$ and $\{\ket{\kdown},\ket{\kpup}\})$, which are time-reversal equivalents. Each pair can, in principle, be used to encode a Kramers qubit.\\ 
 \indent The spin-valley locking of Kramers qubits is expected to possess long spin-valley lifetimes, given that longitudinal energy relaxation requires simultaneous spin flip and momentum relaxation \cite{vanvleck_blg}. On a more fundamental level, time-reversal symmetry prevents spin relaxation via lattice phonons at low magnetic fields, an effect called the \textit{van Vleck cancelation} \cite{vanvleck_orig,vanvleck_blg}. In this context, spin-valley lifetimes of $\approx 30 \ \text{s}$ have recently been measured in single shot spin readout of Kramers qubit in BLG QDs \cite{denisov2024ultralongrelaxationkramersqubit}.  \\
\indent The essential ingredient of the Kramers qubit is SOC \cite{goh2020toward,kane2005quantum}, which is intrinsically weak in the BLG platform (of the order tens of µeV). Recent experiments have therefore focused on TMDC QDs \cite{doi:10.1021/acs.nanolett.3c01779,D3NR03844K,doi:10.1021/acs.nanolett.1c02177,song2015gate,pisoni2018gate}, in which the SOC can reach the order of meV for electrons and even tens of meV for holes. Indeed spin-valley locking is a known property of TMDC monolayers, as typically inferred from optical spectroscopy, where it shows up as circular optical dichroism and spin-valley optical selection rules \cite{valley-dichroism2012}. Evidence of spin-valley locked quantum states for electrons confined to QDs have only recently been reported \cite{doi:10.1021/acs.nanolett.3c01779}. \\
\indent The recent device-level advances have been able to infer critical information on the effective electronic $g$-factors \cite{doi:10.1021/acs.nanolett.3c01779,D3NR03844K,doi:10.1021/acs.nanolett.1c02177,song2015gate,pisoni2018gate} as well as the SOC strengths \cite{doi:10.1021/acs.nanolett.3c01779}. Yet, the assignment to specific spin-valley eigen states and the relative magnitudes of spin and valley $g$-factors have remained uncertain. This work aims to advance a unifying viewpoint on recent experiments by recreating the transport spectroscopy experiments via numerical simulations. We simulate both ground and excited state transport spectroscopy using a numerical experiment platform that utilizes the density matrix rate equations \cite{Beenakker,BM_1,BM_2,Muralidharan_2008,Mukherjee_2023} in the Fock space of a generalized Hamiltonian for 2D-single quantum dots (SQDs) \cite{Mukherjee_2023}. Providing near-exact matches of the transport spectroscopy data from the three distinct experiments lets us distill, on a unifying plane, an overarching set of conclusions on the operating conditions under which spin-valley locking can be achieved. 

\section{Preliminaries\label{sec:preliminaries}}
We start with a generic 2D material platform characterized by a hexagonal Brillouin zone, as shown in  Fig.~\ref{fig:fig_1}(a), over which a single quantum dot (QD) is created and controlled by virtue of voltage-controlled gates \cite{loss1998quantum,hamer2018gate,song2015gate,hanson2007spins}. We describe the QD via the four discrete states shown in Fig.~\ref{fig:fig_1}(b) and develop the Hamiltonian as described in Appendix ~\ref{sec:app_formalism}. Inside the dot, electrons are localized at the $K$ and $\KP$ points with spins $\uparrow$ or $\downarrow$. The intrinsic SOC \cite{banszerus2021spin,PhysRevLett.122.217702,PhysRevB.85.115423,guinea2010spin,island2019spin,PhysRevLett.110.066806,PhysRevB.100.161110} breaks the space inversion symmetry and an external out-of-plane magnetic field breaks the time reversal symmetry, resulting in both spin- and valley-Zeeman effects with their strengths governed by the spin and valley- $g$-factors. In addition, we consider the hopping of electrons between the two valleys \cite{guinea1998spin,morpurgo2006intervalley}, while ignoring any spin-flip processes. The evolution of the energy spectrum as a function of the out-of-plane magnetic field $B_{\perp}$ in Fig.~\ref{fig:fig_1}(c) also shows the regions in which different types of qubits can be operated. \\
\indent The QD model \cite{david2018effective,hensgens2017quantum} depicted in Fig.~\ref{fig:fig_1}(b), comprises four single particle energy levels, with terms $t_v$ and $\Delta_{so}$ accounting for the inter-valley mixing and intrinsic SOC respectively. The $16 \times 16 $ Fock space is number diagonal with five sub-spaces each corresponding to $N=0,1,..,4$ electron(s) in the dot as described in Appendix ~\ref{sec:app_fock_space}. Each block is characterized by a set of eigenstates $\ket{N,i}$ with energy $E_N^i$, where $i$ denotes the $i^\text{th}$ state in the corresponding Fock-subspace.  \\
\indent The abstraction of the transport model is illustrated in Fig.~\ref{fig:fig_1}(c)(Inset). For the results to follow, just like in the experiments, we will simulate transport spectroscopy as a tool to track the transitions, or equivalently, transport channels $ \epsilon_{N}^{ij}=E_N^i-E_{N-1}^j$ \cite{BM_1,BM_2,Mukherjee_2023} between the adjacent $N$-particle Fock spaces. Depending on the experiments we simulate, the experimental data is mimicked by tracking the ground state transport channel 
\begin{equation}
    \epsilon_{N}^{00}=E_N^0-E_{N-1}^0,
\end{equation}
with the applied magnetic field, or, the translation of the excited state transport channel as a function of the  applied magnetic field,
\begin{equation}
    \epsilon_N^{Z_{ij}}=\epsilon_N^{i0}-\epsilon_N^{j0},
\end{equation}
in the case of excited state spectroscopy. 
Thus, depending on the experiment we are trying to simulate, the conclusions are drawn based on tracking either $V_{SD}$ or $V_G$, while maintaining the other at a constant value for the corresponding transition, as depicted via the horizontal or vertical eye-guides in Fig.~\ref{fig:fig_1}(d).
\begin{figure*}
\includegraphics[width=0.7\textwidth]{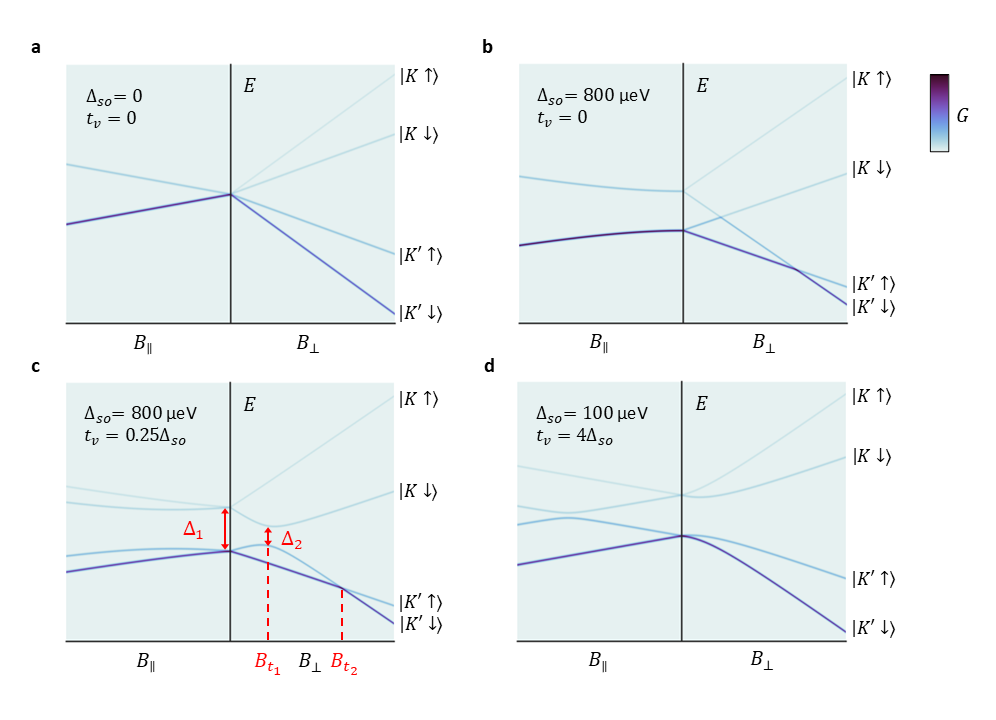}
\captionsetup{justification=raggedright,singlelinecheck=false}
\caption[width=\textwidth]{
Evolution of the energy states of 2D QDs with applied magnetic fields. Variation of the single particle spectrum at the $K$ and $\KP$ points as a function of $B_{\parallel}$ and $B_{\perp}$. \textbf{(a)} The variation of energy levels when both SOC parameter $\Delta_{so}$ and intervalley mixing $t_v$ play no role, \textbf{(b)} when only $\Delta_{so}$ is present, \textbf{(c)} with $t_v$ also included, where two threshold fields $B_{t_1}$ and $B_{t_2}$ define operating regions of spin-valley qubits, valley qubits and spin qubits respectively (to be discussed later), and \textbf{(d)} when $t_v$ is dominant in comparison with $\Delta_{so}$. 
} 
\label{fig:fig_2}
\end{figure*}

\begin{figure*}

\centering
\includegraphics[width=0.7\textwidth]{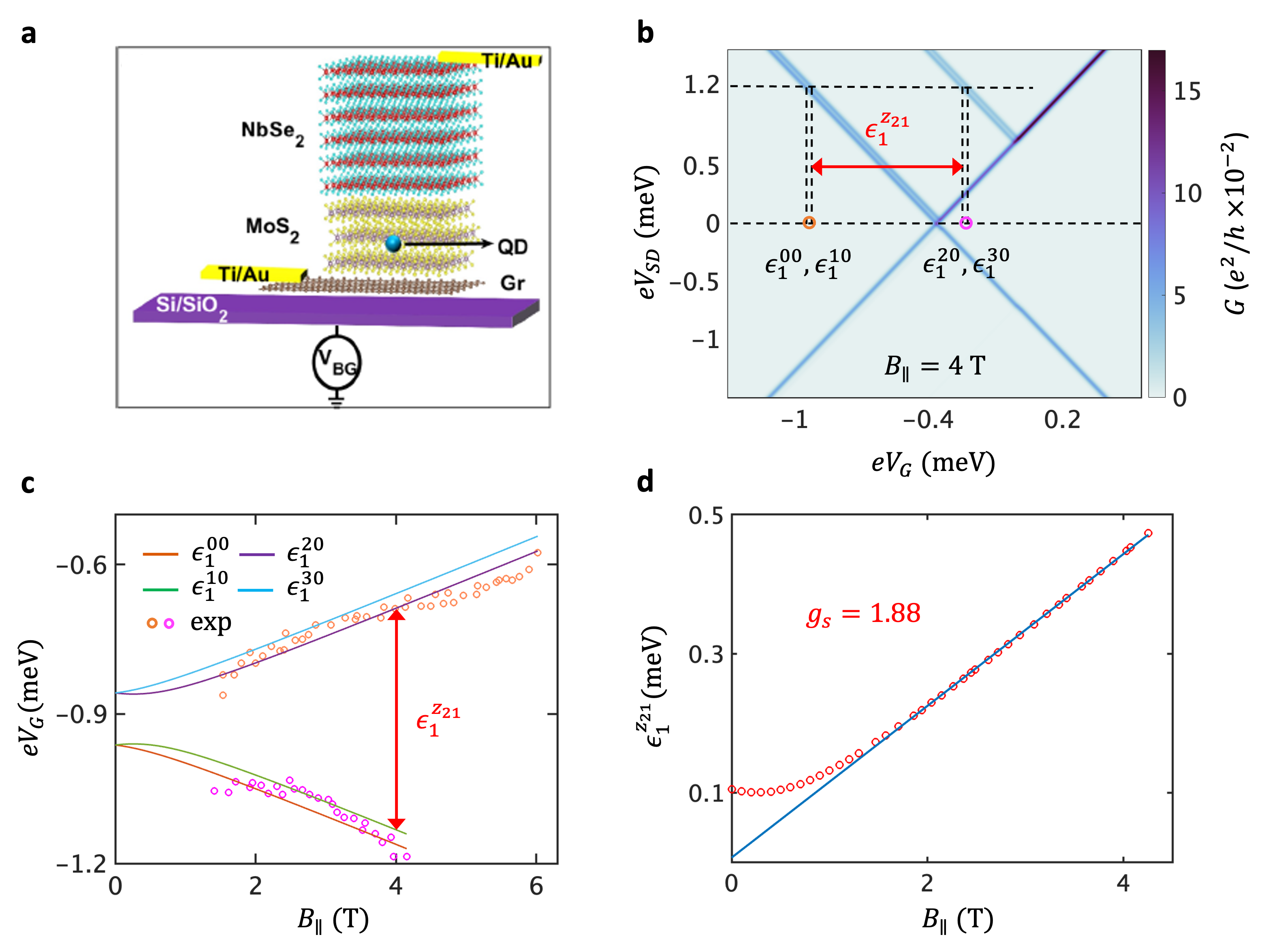}

\captionsetup{justification=raggedright,singlelinecheck=false}
\caption[width=\textwidth]{
Interpreting the experimental data from Devidas \textit{et al.}, (2021) \cite{doi:10.1021/acs.nanolett.1c02177}, from the excited state spectroscopy of an atomic point defect QD in MoS2 with superconducting drain electrodes. \textbf{(a)} Schematic of the device (reproduced from \cite{doi:10.1021/acs.nanolett.1c02177}). \textbf{(b)} Coulomb diamond simulations corresponding to the $N=0 \leftrightarrow N=1$ transition with unequal left and right tunneling rates showing Zeeman splitting in the presence of an in-plane magnetic field. The horizontal eye-guides sketched show the extraction process. \textbf{(c)} Simulation fits of the one electron energy states as a function of $B_{\parallel}$. This gives us an estimate that $\Delta_{so}=100$ µeV and $t_v=15$ µeV. \textbf{(d)} Fitting a straight line to the $\epsilon_1^{Z_{21}}$ extracted from (\textbf{(c)}) gives us a spin $g$-factor $g_s=1.88$, verifying the accuracy of our fit.
} 
\label{fig:fig_3}
\end{figure*}
The QD Hamiltonian is based on the following experimental estimates: a) the size of each lateral quantum dot typically lies in the range of $20-130$ nm \cite{goh2020toward,pawlowski2024single,zhang2017electrotunable}, b) the dielectric constant for MoS$_2$ is in the range of $\epsilon_r=4-20$ \cite{santos2013electrically,li2016charge} depending on the number of layers. Thus, the corresponding Coulomb repulsion energy ($U_C$) lies in the range $2-16$ meV \cite{david2018effective,pawlowski2024single,wang2018electrical,pisoni2018gate}. Complying with these values, we set $U_{C}=5$ meV. The spin-orbit parameter $\Delta_{so}$ is expected to be in the range of $100-3000$ µeV and the $g$-factors to be $g_s \approx 2$ and $g_v \approx 0.5-6.5$ \cite{kormanyos2014spin,marinov2017resolving,doi:10.1021/acs.nanolett.3c01779,kosmider2013large,kormanyos2015landau,pawlowski2019spin,macneill2015breaking}. Therefore, for our model we consider 
$g_s=2$ and $g_v=6$ for MoS$_2$. Setting the coupling rates with the contacts as $\gamma_L=\gamma_R=5\times10^{-6}$, we simulate the current under different vectorized magnetic field as described in Appendix ~\ref{app_transport}. Fig.~\ref{fig:fig_2} shows the ground and excited states for $N=1$ for different $\Delta_{so}$ and $t_v$ as generated from the simulated excited state magneto-spectroscopy.\\ 
\indent A comparison between Fig.~\ref{fig:fig_2}(a) and~\ref{fig:fig_2}(b) confirms the fact that the SOC leads to splitting of the four degenerate states into pair of Kramers doublets. The inter-valley mixing term, $t_v$, whose magnitude is typically a fraction of $\Delta_{so}$ and  could be impurity assisted \cite{szechenyi2018impurity}, splits the pair of Kramers doublets into four states in the presence of $B_{\parallel}$ (compare Fig.~\ref{fig:fig_2}b and c). Further, $t_v$ results in the anti crossing, between the two valleys with identical spins with increasing $B_{\perp}$. This anti crossing of energy states with $B_{\perp}$ occurs at $B_{t_1}$, where
\begin{equation}
B_{t_1} = \frac{\Delta_{so}}{g_v\mu_B},
\label{eq:B_t1}
\end{equation}
and the energy gap at the anti crossing is $\Delta_2$, where 
\begin{equation}
    \Delta_2 = 2t_v
\end{equation}

Further there is a crossing of the two spin states from the same valley at $B_{t_2}$, where
\begin{equation}
 B_{t_2} = \frac{1}{\mu_B}\sqrt{\frac{\Delta_{so}^2}{g_s^2} - \frac{4t_v^2}{g_v^2 - g_s^2}}.
\label{eq:B_t2}
\end{equation}

The energy separation $(\Delta_1)$ at  $B=0$ between the pair of Kramers doublets depends on the intrinsic spin orbit coupling and the inter-valley mixing term and is given by 

\begin{equation}
    \Delta_1 = \sqrt{\Delta_{so}^2 + \Delta_{2}^2}.
    \label{eq:delta}
\end{equation}

The slopes of the four $N=1$ eigenstates when $B_\parallel=0$ and 0 < B < $B_{t_1}$, neglecting the second order terms in $t_v$, are 
\begin{subequations}
\begin{align}
\frac{dE_1^{0(3)}}{dB_{\perp}} &= \frac{\mu_B}{2}(g_s \pm g_v)\\
\frac{dE_1^{1(2)}}{dB_{\perp}} &= \frac{\mu_B}{2}(-g_s \mp g_v)
\end{align}
\label{eq:perp_slopes}
\end{subequations}
Thus, in principle, we could estimate the parameters $\Delta_{so}$, $t_v$, $g_s$ and $g_v$, if we are able to extract any of the quantities in ~\eqref{eq:B_t1} to ~\eqref{eq:perp_slopes} from a transport spectroscopy experiment. The remaining quantities can then also help us in independently verifying our estimates. Moreover, $\Delta_{so}$, $t_v$ and $g_s$ can also be obtained by fitting the $N=1$ eigenvalue equations for the case $B_\perp=0$ as given in ~\eqref{eq:n1par}

\section{Simulating the Experiments \label{sec:formalism}}
 So far only three experiments on TMDC QDs have reported results from transport spectroscopy of few electron spin states \cite{doi:10.1021/acs.nanolett.1c02177,D3NR03844K,doi:10.1021/acs.nanolett.3c01779}, which we now reproduce one by one.\\
 
\begin{figure*}

\centering
\includegraphics[width=0.9\textwidth]{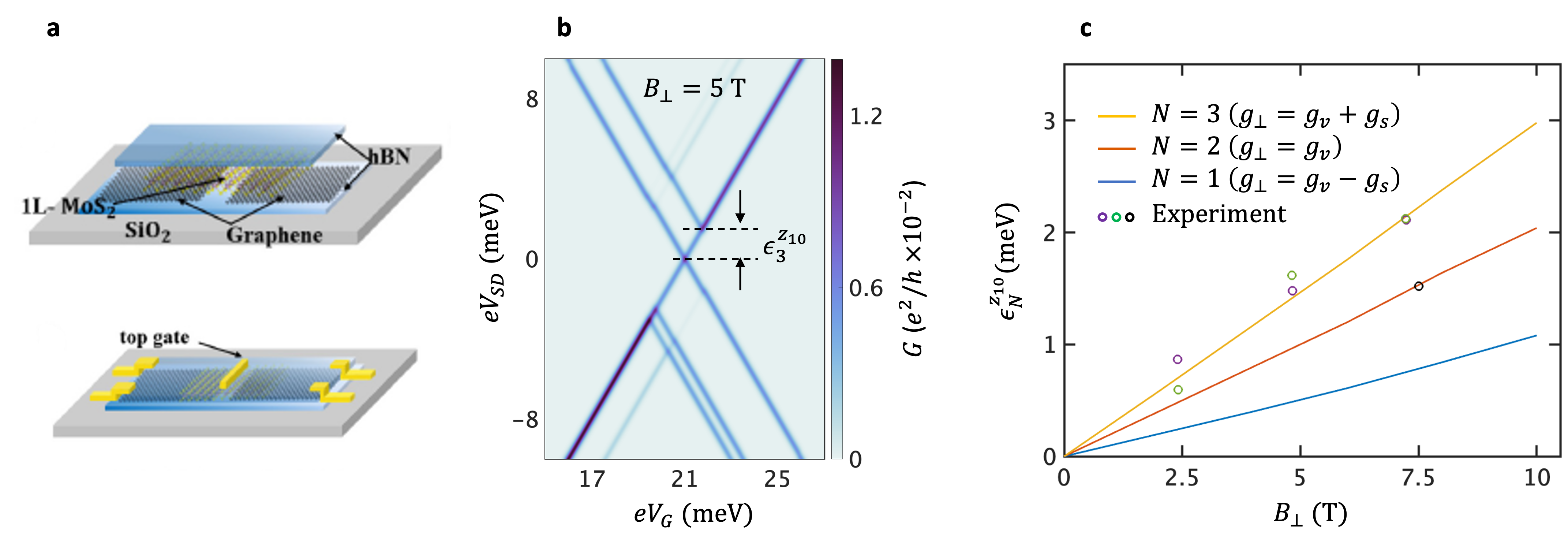}

\captionsetup{justification=raggedright,singlelinecheck=false}
\caption[width=\textwidth]{
Interpreting the experimental data from Kumar \textit{et al.}, (2023) \cite{D3NR03844K}, from excited state spectroscopy of a monolayer MoS$_2$ quantum dot. \textbf{(a)} Schematic of the device (reproduced from \cite{D3NR03844K}). \textbf{(b)} Coulomb diamond simulations corresponding to the $N=2 \leftrightarrow N=3$ transitions with unequal left and right tunneling rates showing Zeeman splitting in the presence of an out-of-plane magnetic field. It also depicts the extraction of experimental data. \textbf{(c)} Fitting of the experimental Zeeman splitting data with $B_{\perp}$ for diamonds corresponding to different electron numbers. This gives us an estimate of $g_v=3.5$ \& $g_s=1.67$ and the maximum magnitude of $B_{\perp}$ gives us $\Delta_{so}\ge 1.52$ meV.
}
\label{fig:fig_4}
\end{figure*}
\indent We start with the experimental results reported by Devidas \textit{et al.}, (2021) \cite{doi:10.1021/acs.nanolett.1c02177} performed on atomic point defect QD in MoS$_2$ with superconducting drain electrodes, subjected to an in-plane magnetic field ($B_{\parallel}$). The device geometry is shown in Fig.~\ref{fig:fig_3}(a). Fig.~\ref{fig:fig_3}(b) elucidates the method used by the authors to estimate the ordinate of Fig.~\ref{fig:fig_3}(c). Furthermore, the experimental data (dots) in Fig.~\ref{fig:fig_3}(c) was graphically extracted from the paper (reported only for the region $B_{\parallel} > 1.5 \ \text{T}$), and used to extract $\Delta_{so}=100$ µeV and $t_v=15$ µeV, by fitting the data to the four $N=1$ eigenvalue equations given in ~\eqref{eq:n1par} (Fig.~\ref{fig:fig_3}(c)). From the slope of $\epsilon_1^{Z_{21}}=E_1^2-E_1^1$, we further extract $g_s=1.88$ in Fig.~\ref{fig:fig_3}(d), which agrees well with the reported value. We are not able to make any estimates of $g_v$ as there is no spectroscopy data reported with varying $B_{\perp}$. \\

\indent We now turn our attention to the second experiment by Kumar \textit{et al.}, (2023) \cite{D3NR03844K}, featuring the excited state spectroscopy ($\epsilon_N^{Z_{10}}$) in a gate defined QD in a monolayer MoS$_2$, as a function of $B_{\perp}$, as depicted in Fig.~\ref{fig:fig_4}. They have reported two different values of effective out-of-plane $g$-factor, namely $g_\perp=5.2$ and $g_\perp=3.5$, for excited states corresponding to different Coulomb diamonds. While the authors believe that is an artefact of point defects, we believe that our model explains it through a consistent set of spin and valley $g$-factors that result in different effective out-of-plane $g$-factors simply due to the varying occupation numbers in the Coulomb diamonds. Assuming $B_\perp < B_{t_1}$ (compare Fig.~\ref{fig:fig_2}(c)), and from equations ~\eqref{eq:n1perp}, ~\eqref{eq:n2perp3} and ~\eqref{eq:n3perp}, we have 
\begin{align}
     \epsilon_1^{Z_{10}}&=E_1^1-E_1^0 \nonumber & (N=1) \\ 
    &=(g_v-g_s)\mu_B B_\perp,
\end{align}

\begin{align}
    \epsilon_2^{Z_{10}}&=E_2^1-E_2^0 = E_1^2-E_1^1 \nonumber  & (N=2)\\
    &=\Delta_{so}-g_v\mu_B B_\perp,
\end{align}

\begin{align}
    \epsilon_3^{Z_{10}}&=E_3^1-E_3^0 = E_1^3-E_1^2 \nonumber  & (N=3)\\
    &=(g_v+g_s)\mu_B B_\perp.
\end{align}
Defining $\epsilon_N^{Z_{10}}=g_\perp \mu_B B_\perp$, we can see that the effective $g_\perp$ depends on both the electron number in the dot and the strength of $B_\perp$ in relation to $\Delta_{so}$, as it is a combination of spin and valley $g$-factors as summarized in Tab.~\ref{tab:slopes}. 
\begin{table}[h!]
    \centering
    \caption{Dependence of the slope of $\epsilon_N^{Z_{10}}$, i.e., the effective $g_\perp$, on the number of electrons in the dot and the region where $B_\perp$ is operating.}
    \begin{tabular}{|c|c|c|c|}
    \hline\hline
     & $N=1$ & $N=2$  & $N=3$\\
    \hline
    $B_\perp<B_{t_1}$ & $g_v-g_s$ & $g_v$ & $g_v+g_s$ \\
    $B_{t_1}<B_\perp<B_{t_2}$ & $g_s$ & $g_v$ & $g_s$ \\
    $B_\perp>B_{t_2}$ & $g_s$ & $g_v-g_s$ & $g_s$ \\
    \hline\hline
    \end{tabular}
    \label{tab:slopes}
\end{table}
\begin{table}[h!]
    \centering
    \caption{Dependence of the offset of the shift in excited state at $B_\perp=0$.}
    \begin{tabular}{|c|c|c|c|}
    \hline\hline
     & $N=1$ & $N=2$  & $N=3$ \\
    \hline
     $\epsilon_N^{Z_{10}}$ at $B_\perp=0$ & 0 & $\Delta_{so}$ & 0 \\
    \hline\hline
    \end{tabular}
    \label{tab:offsets}
\end{table}

\begin{figure*}

\centering
\includegraphics[width=1\textwidth]{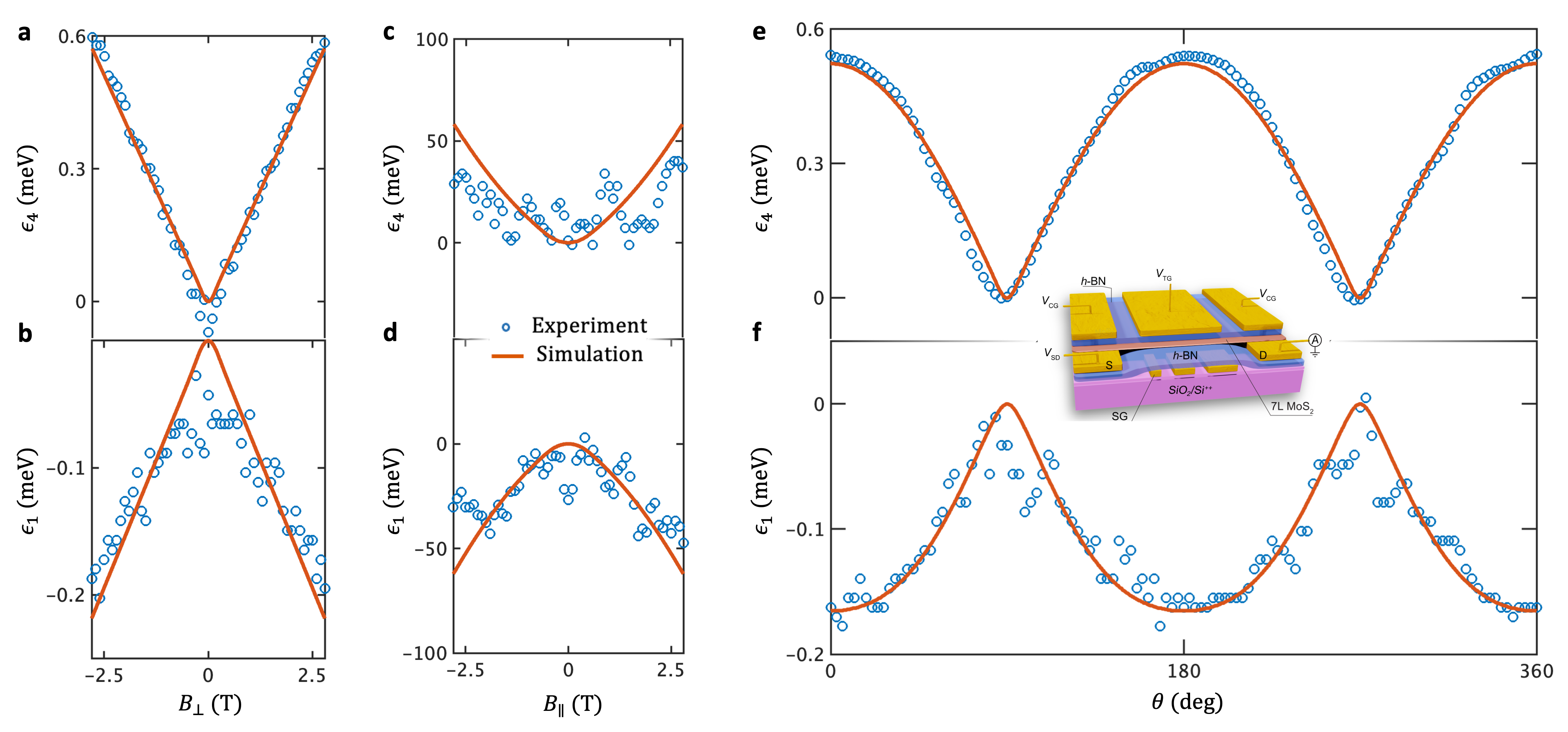}

\captionsetup{justification=raggedright,singlelinecheck=false}
\caption[width=\textwidth]{
Interpreting the experimental data from Krishnan \textit{et al.}, (2023) \cite{doi:10.1021/acs.nanolett.3c01779}, from ground state spectroscopy of a QD in 7-layer MoS$_2$. \textbf{(a) and (b).}\footnote{The data has been updated from the original manuscript by the original authors as there was an error  in the gate lever used. } Extraction of $g_s = 2.2$ and  $g_v=5$ based on fits to experimental data with respect to $B_{\perp}$. \textbf{(c) and (d).} Extraction of $\Delta_{so} = 852$ µeV and $t_v = 65$ µeV based on fits to experimental data with respect to $B_{\parallel}$. \textbf{(e) and (f).} Closing the loop to demonstrate the consistency of the parameter extraction process. Fit of an independent experimental data set that features an anisotropic magnetic field with |B|=2.8T with varying angle (inset - schematic of the device reproduced from \cite{doi:10.1021/acs.nanolett.3c01779}).
} 
\label{fig:fig_5}
\end{figure*}

We know from the reported experimental data that the offset at $B=0$ for the case $g_\perp=5.2$ is zero. Based on Tab.~\ref{tab:offsets}, we can say that the case $g_\perp=5.2$ corresponds to an electron occupation of $N=1$ or $N=3$. However, since the other reported $g_{\perp}=3.5$ is a smaller slope, we can now conclude that the electron occupation number is $N=2$ and $N=3$ for $g_\perp=3.5$ and $g_\perp=5.2$ respectively (first row of Tab.~\ref{tab:slopes}). From this, we can conclude that $g_v=3.5$ and $g_s=1.7$ for the sample, which agrees very well with what is expected for MoS$_2$. Further, as the slopes in Fig.~\ref{fig:fig_4}(c) do not change until $B_{\perp}=7.5$ T, $B_{t_1}$ must be at least $7.5$ T, which in turn enforces a lower bound on the SOC strength, i.e. $\Delta_{so}\ge1.52$ meV based on ~\eqref{eq:B_t1}. We are not able to make any estimates of $t_v$ as there is no spectroscopy data reported with applied in-plane field. \\
\indent Finally, Krishnan \textit{et al.}, (2023) \cite{doi:10.1021/acs.nanolett.3c01779}, recently reported a detailed ground state magneto-spectroscopy in which the Zeeman anisotropy of few electron spin-valley states was mapped out in a 7-layer MoS$_2$ transistor (inset in Fig.~\ref{fig:fig_5}(e) \&~\ref{fig:fig_5}(f)) in vectorized (both in-plane and out-of-plane) magnetic fields. From the detailed analysis of two ground state transitions (CB peaks), and their shifts with $B(\theta,l)$ (reproduced in Fig.~\ref{fig:fig_5}), they were able to confirm spin-valley locking for the first time at the few electron level, and extract effective $g$-factors. To fit the data, we define $\epsilon_N$ such that

\begin{equation}
    \epsilon_N(B) = \epsilon_N^{00}(B) - \epsilon_N^{00}(B=0).
    \label{epsilon N}
\end{equation}

\indent By comparing with Fig.~\ref{fig:fig_2}(c), we can conclude that the two ground state transition in Fig.~\ref{fig:fig_5}(a) \&~\ref{fig:fig_5}(b) most likely corresponds to $\epsilon_1$ and $\epsilon_4$. From the slopes defined in ~\eqref{eq:perp_slopes}, we extract $g_s=2.2$ and $g_v=5$ from Fig.~\ref{fig:fig_5}(a) \&~\ref{fig:fig_5}(b). This is in good agreement with previous works such as Kumar \textit{et al.}, (2023) \cite{D3NR03844K}, albeit slightly larger spin and valley $g$-factors. The maximum magnitude of the $B$ field at $2.8 \ \text{T}$ limits us from using the expressions for $B_{t_1}$ and $B_{t_2}$. However, by fitting the $N=1$ eigenvalue equations (~\eqref{eq:n1par}) to the variation of $\epsilon_1$ and $\epsilon_4$ with $B_\parallel$ to the data in Fig.~\ref{fig:fig_5}(c) and (d), we obtain an estimate of $\Delta_{so}=852 \ \text{µeV}$ and $t_v=65 \ \text{µeV}$. Based on these extracted values of $\Delta_{so}$, $t_v$, $g_s$, and $g_v$, we predict the ground state transitions ($\epsilon_1$ and $\epsilon_4$) with the angle variation in the magnetic field sweep at a fixed magnitude of $2.8$T and compare with the experimental results in Fig. ~\ref{fig:fig_5}(e) and (f). This aspect also ascertains the consistency in the parameter extraction process. We are also able to ascertain that QDs formed in MoS$_2$ with up to a few layers may also prove suitable for implementing a Kramers qubit, owing to the significant value of the SOC strength.\\
\indent This is the only experiment amongst the three where we are able to extract all the four parameters (Tab.~\ref{tab:summary}), as  Krishnan \textit{et al.} have reported spectroscopy data for both in-plane and out-of-plane magnetic fields while Devidas \textit{et al.} and Kumar \textit{et al.} have reported spectroscopy data only in in-plane and only in out-of-plane magnetic fields respectively. Hence, in order to isolate and extract all the four parameters for TMDCs, experiments need to record spectroscopy data for both in-plane and out-of-plane fields given $B_{t_1}$ may be larger than the operating range of the experimental setup.   \\
\indent To summarize, a salient outcome of our analysis is the ability to extract a consistent set of spin and valley $g$-factors for MoS$_2$ (Tab.~\ref{tab:summary}), and distinguish each individual contribution ($g_s$ and $g_v$) to the effective transport $g$-factors so far reported.  In addition, our work also unifies the extraction  of $\Delta_{so}$ across three different QDs realised in MoS$_2$ crystals of different number of layers, which confirms that $\Delta_{so}$ indeed decreases with layer number (left to right in Tab.~\ref{tab:summary}) as expected ~\cite{chang2014thickness}. Our results also shed light on the existence of inter-valley mixing ratified via the analysis of transport spectroscopy experiments, whenever, an out-of-plane magnetic field is applied. 

\begin{table*}[t]
    \centering
    \caption{Summary of all the extractions.}
    \begin{tabular}{|c|c|c|c|}
    \hline\hline
     & \textbf{Kumar \textit{et al}.}  
     & \textbf{Krishnan \textit{et al}.}
     & \textbf{Devidas \textit{et al}.} \\
    \hline
    No. of MoS$_2$ Layers & 1 \footnote{Based on photo-luminescence, TEM and AFM measurements} & 7 \footnote{Based on Raman spectrum and AFM measurements} & $>$ 7 \footnote{The number of layers is not mentioned in the manuscript. The fabrication process does not intend to minimise layers.}  \\
    Electrostatic Environment & dielectric hBN & dielectric hBN & Heterostructure  \\
    Contacts & Graphene & Ti/Au (Gated) & Graphene and NbSe$_2$\\
    Sample Temperature & 1 K & 30mK \footnote{Authors reported an electron temperature of 150mK} & 30mK \\
    Maximum $|B|$ & 7.5T & 2.8T & 6T \\
    Orientation of B 
        & $B_{\perp}$ only 
        & $B_{\perp}$, $B_{\parallel}$, and $\theta$ at fixed $|B|$ 
        & $B_{\parallel}$ only \\
    Magneto-spectroscopy measurement 
        & Excited state 
        & Ground state  
        & Excited state \\
     $\Delta_{so}$ (µeV) & $>$1520 & 852 & 100  \\ 
     $t_v$ (µeV)  & Insufficient data & 65 & 15\\
     $g_s$  & 1.7 & 2.2 & 1.88 \\
     $g_v$ & 3.5 & 5 & Insufficient data  \\
     \hline\hline
    \end{tabular}
    \label{tab:summary}
\end{table*}

\section{Further outlook\label{sec:insights}}

\begin{table}[h]
    \centering
    \caption{Comparing the MoS$_2$ and the BLG plaforms for the Kramers and valley qubits.}
    \begin{tabular}{|c|c|c|}
    \hline\hline
     & \textbf{MoS$_2$ \footnote{Krishnan \textit{et al}. \cite{doi:10.1021/acs.nanolett.3c01779}}}
     & \textbf{BLG \footnote{Denisov \textit{et al}. \cite{denisov2024ultralongrelaxationkramersqubit}}} \\
    \hline
    $\Delta_{so}$ (µeV)  & 852 & 64  \\
    $t_v$ (µeV)  & 65  & 5   \\
    $g_s$  & 2.2 & 2  \\
    $g_v$  & 5  & 14.5  \\
    \hline
    \multicolumn{3}{|c|}{Kramer qubits for out-of-plane $|B| < B_{t_1}$} \\
     \hline
    $B_{t_1}$ & 2.9 T & 76 mT \\
    Energy gap between the two states (µeV) & 415 & 46 \\
    Equivalent temperature for the Energy Gap & 4.8 K & 534 mK \\
 \hline
 \multicolumn{3}{|c|}{Valley qubits for out-of-plane $|B| \approx  B_{t_1}$} \\
     \hline
    Range of |B| around $B_{t_1}$ & 440 mT & 3.5 mT \\
    Energy gap between the two states (µeV) & 130 & 10 \\
    Equivalent temperature for the Energy Gap & 1.5 K & 116 mK \\
 \hline
     \hline\hline
    \end{tabular}
    \label{tab:mos2_blg}
\end{table}

\begin{figure}[h]
\centering
\includegraphics[width=0.9\columnwidth]{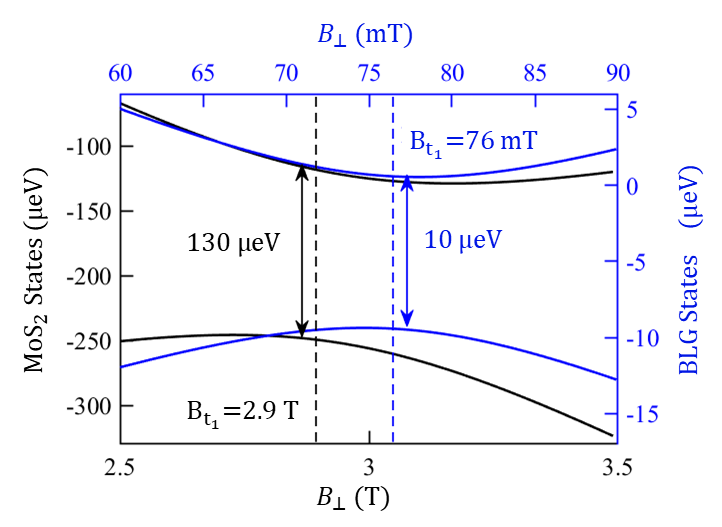}
\caption{\label{fig:fig_6} Analyzing the valley anti-crossing and establishing the strength of spin-valley locking. The $K \downarrow$ and $\KP\downarrow$ states with varying $B_{\perp}$ for the BLG (blue axes and lines) and the MoS$_2$ (black axes and lines) platforms. The value of $B_{t_1}$ for both cases are indicated on respective axes and color at the anti-crossing point where the gap between the two states is minimum. The average energy gap in the anti-crossing region is indicated for BLG (blue) and for MoS$_2$ (black).
} 

\end{figure}
 Having drawn a set of unifying conclusions, we can further develop testable hypotheses that may serve as a guide 
 to experiments in the future. In this context, let us now use our results at hand to make a cross-platform outlook for the realization of Kramers qubit and the valley qubit.\\
\indent The stability of a Kramers pair $({\ket{\kup}},{\ket{\kpdown}}$) determines the degree of spin-valley locking. At very low temperatures, the “locking” can be lifted primarily by increasing the out-of-plane magnetic field. We can use this as a possible quantifier for spin-valley locking. Specifically, spin-valley locking can sustain up to an energy scale of $\Delta_{so}$. The out-of-plane magnetic field required to lift it is given by \eqref{eq:B_t1} and is a function of both $\Delta_{so}$ and $g_v$. Based on this, we can make a preliminary assessment that the MoS$_2$ platform has one to two orders of magnitude stronger spin-valley locking compared to the BLG platform, as seen from $B_{t_1}$ in Fig.~\ref{fig:fig_6}, which is, 2.9 T for MoS$_2$ in comparison to 76 mT for the BLG system (also summarized in Tab.~\ref{tab:mos2_blg}). Additionally, as a consequence of a larger $\Delta_{so}$, the MoS$_2$ platform can sustain a larger intervalley mixing $t_v$, since a larger $t_v$ can also lift spin-valley locking as clearly depicted in Fig.~\ref{fig:fig_2}(d). While the intrinsic gap in BLG is < $100 \ \text{µeV}$, efforts are made to induce SOC into graphene by proximity, e.g. in heterostructures with TMDC crystals ~\cite{proximity1_PhysRevB.100.085412,proximity2_PhysRevB.106.125417,proximity3_PhysRevB.108.235166}. Moreover, a larger energy gap can also make conventional forms of single-shot projective measurements based on the Elzerman \cite{Elzerman2004} or the Morello \cite{Morello2010} techniques more reliable. \\
\indent In terms of the operation of valley qubits, our simulations also project that there is a region around $B_{t_1}$ where the superpositions of $\ket{K \downarrow}$ and $\ket{\KP\downarrow}$ exist with a nearly constant energy gap, as depicted in Fig.~\ref{fig:fig_6} and summarized in Tab.~\ref{tab:mos2_blg}. Once again, this energy gap is an order of magnitude larger in MoS$_2$ ($\approx130 \ \text{µeV}$), while in the case of the BLG platform, it is practically undetectable. This has also been observed in various experiments that measure relaxation times in the BLG QD \cite{valley_relax_bukkar,PRXQuantum.3.020343,exp_valley_relax_bukkard, denisov2024ultralongrelaxationkramersqubit}. However, experiments have also measured valley relaxation times in BLG double quantum dots ~\cite{valley_dqd_blg2024}. The larger energy gap at the anti-crossing makes it more feasible to operate valley qubits in MoS$_2$ QD, where the superposition of the valley states can, in principle, be controlled by simply changing the strength of the out-of-plane magnetic field. We can hence affirmatively conclude a TMDC QD can be used to detect spin-valley and valley relaxation times through the single-shot Elzerman readout around the valley anti-crossing point. \\

\section{Conclusions \label{sec:conclusion}}
We have developed a unified analytical and numerical framework to reverse engineer data from transport spectrosopy experiments and view them on a unified footing. This helps in designing transport spectroscopy experiments in 2D semiconductor quantum dots using an exhaustive set of parameters capturing the delicate interplay of Coulomb interactions, spin and valley Zeeman effects, intrinsic SOC and inter-valley mixing. Analyzing the relevant Fock-subspaces of the generalized Hamiltonian, coupled with the density matrix master equation technique for transport, we have been able to extract critical transport parameters thus unifying the findings of recent experiments on spin-valley states in MoS$_2$ single quantum dots. While providing an encompassing framework, our work also stresses the importance of simulating experimental data to design new experiments. This may aid to conclusively measure valley-relaxation times using single-shot projective readout methods in TMDC quantum dots for the first time. Although our unified framework is currently designed for transport simulations, it is easily expanded to simulate spin-valley lifetime experiments, the next milestone in TMDC QD research.

\section*{Acknowledgements} The authors wish to acknowledge Guido Burkard and Andr\'as P\'alyi for insightful discussions. The author BM acknowledges funding from the Science and Engineering Research Board of India, through Grant No. MTR/2021/000388, under the MATRICS scheme. 
The authors BM and SK acknowledge funding from the Dhananjay Joshi Endowment award from IIT Bombay and the Inani Chair Professorship fund. The author SK acknowledges the Swasth Research Fellowship. The author BW acknowledges the support of the National Research Foundation (NRF) Singapore, under the Competitive Research Program “Towards On-Chip Topological Quantum Devices” (NRF-CRP21-2018-0001), with further support from the Singapore Ministry of Education (MOE) Academic Research Fund Tier 3 grant (MOE-MOET32023-0003) “Quantum Geometric Advantage” and the Air Force Office of Scientific Research under award number FA2386-24-1-4064.

\appendix

\section{Formalism\label{sec:app_formalism}}
    A schematic of the setup is shown in Fig.~\ref{fig:fig_1}(c)(Inset) of the main text. The device consists of a single quantum dot defined and manipulated using voltage controlled gates. Besides the onsite energy, there is an onsite Coulomb repulsion, with energy $U_{C}$, between each pair of electrons in the dot. Inside the dot, the conduction electrons are localized on either of the two valleys: $K$ or $\KP$, and can have a spin $\uparrow$ or spin $\downarrow$. Thus, for a low energy single dot, there are $4$ states available, namely, $\ket{\kup},\ket{\kdown},\ket{\kpup},\ket{\kpdown}$. In the presence of an electric field from the voltage controlled gates, the space inversion symmetry is broken, leading to the splitting of the four energy states into two Kramer pairs through an intrinsic spin-orbit (SO) coupling \cite{banszerus2021spin,PhysRevLett.122.217702,PhysRevB.85.115423,guinea2010spin,island2019spin,PhysRevLett.110.066806,PhysRevB.100.161110}. The energy of the pair $\{\ket{\kup},\ket{\kpdown}\}$ is increased by an amount $\frac{1}{2}\Delta_{so}$, while that of the pair $\{\ket{\kdown},\ket{\kpup}\}$ is decreased by the same amount. Subsequently, in the presence of an external magnetic field the time inversion symmetry is also broken, thus the degeneracy of the states in each of the Kramer pairs is further lifted. This effect is classified into the spin-Zeeman and the valley-Zeeman effects. In the case of an out-of-plane magnetic field $(B_\perp)$, the energy shift due to the spin-Zeeman splitting is given by $\sigma g_s\mu_BB_\perp$, while that due to the valley-Zeeman splitting is given by $\tau g_v\mu_BB_\perp$. Whereas in the presence of a parallel magnetic field $(B_x,B_y)$ only the spin states split into two with their energy difference given by $\sigma g_s\mu_B(B_x-iB_y)$, or equivalently, $\sigma g_s\mu_BB_\parallel$. Here, $\sigma$ is the spin ($\sigma=+\frac{1}{2}$ for spin $\uparrow$, and $\sigma=-\frac{1}{2}$ for spin $\downarrow$) and $\tau$ is the valley pseudo-spin ($\tau=+\frac{1}{2}$ for valley $K$, and $\tau=-\frac{1}{2}$ for valley $\KP$). $\mu_B=5.79\times10^{-5}$ eVT$^{-1}$ is the Bohr magneton, and $g_s$ \& $g_v$ are the spin and valley g-factors respectively. In addition to splitting of the states, an electron can also jump from one of the four states to any of the other three states due to external disturbances in the form of scatterers, magnetic impurities, nuclear spin, time dependent external magnetic fields, etc \cite{guinea1998spin,morpurgo2006intervalley}. These are represented by intervalley mixing $(t_v)$, spin flipping $(t_s)$, and finally the simultaneous valley and spin flipping $(t_{vs})$. Since TMDCs have isotopes with zero nuclear spin, and we only consider the case where there are no magnetic impurities in the sample, as is most often the case in experiments, we can safely ignore the effects of spin flipping. Based on these considerations the Hamiltonian takes the form

\begingroup
\allowdisplaybreaks
\begin{align}
    \hat{H}_{\text{SQD}} &= \underbrace{\epsilon\hat{n}}_{\text{Onsite energy}}+\underbrace{\frac{U_{C}}{2}(\hat{n}^2-\hat{n})}_{\text{Onsite repulsion}}\nonumber\\
    &+\underbrace{\frac{\Delta_{so}}{2}\sum_{\tau,\sigma}\hat{c}^\dagger_{\sigma\tau}(\pmb{\sigma}_3)_{\sigma\sigma}(\pmb{\tau}_3)_{\tau\tau}\hat{c}_{\sigma\tau}}_{\text{Spin orbit coupling}}\nonumber\\
    &+ \underbrace{\frac{\mu_BB_\perp}{2}\bigg(g_s\sum_{\tau,\sigma}\hat{c}^\dagger_{\sigma\tau}(\pmb{\sigma}_3)_{\sigma\sigma}\hat{c}_{\sigma\tau} + g_v\sum_{\tau,\sigma}\hat{c}^\dagger_{\sigma\tau}(\pmb{\tau}_3)_{\tau\tau}\hat{c}_{\sigma\tau}\bigg)}_{\text{Perpendicular magnetic field}}\nonumber\\
    &+ \underbrace{\frac{\mu_Bg_s}{2}\bigg(B_\parallel\sum_\tau\hat{c}^\dagger_{\uparrow\tau}\hat{c}_{\downarrow\tau}+\text{h.c.}\bigg)}_{\text{Parallel magnetic field}}\nonumber\\
    &+ \underbrace{t_v\big(\sum_\sigma\hat{c}^\dagger_{\sigma K}\hat{c}_{\sigma K'}+\text{h.c.}\big)}_{\text{Intervalley scattering}}
    \label{eq:fermi_hubbard_hamiltonian}
\end{align} 
\endgroup
%
where the summations are defined over $\sigma\in\{\uparrow,\downarrow\}$ and $\tau\in\{K,\KP\}$. The terms $\bm{\sigma}_3$ ($\bm{\tau}_3$) is the z-component of the Pauli matrix for the spin(valley pseudo-spin), defined as 
\begin{subequations}
\begin{align}
    \left(\bm{\sigma}_3\right)_{\uparrow\uparrow}=1;& \left(\bm{\sigma}_3\right)_{\downarrow\downarrow}=-1\\
    \left(\bm{\sigma}_3\right)_{\uparrow\downarrow} =0;& \left(\bm{\sigma}_3\right)_{\downarrow\uparrow}=0;\\
    \left(\bm{\tau}_3\right)_{KK}=1;&
    \left(\bm{\tau}_3\right)_{\KP\KP}=-1;\\
    \left(\bm{\tau}_3\right)_{K\KP}=0;&
    \left(\bm{\tau}_3\right)_{\KP K}=0;
\end{align}
\end{subequations}
The symbol $\hat{n}$ denotes the number operator for the number of electrons in the dot, formulated as
\begin{align}
    \hat{n}=\sum_{\sigma,\tau}\hat{c}^\dagger_{\sigma\tau}\hat{c}_{\sigma\tau}
\end{align}
where $\hat{c}^{(\dagger)}_{\sigma\tau}$ is the annihilation (creation) operator for an electron in the dot with spin $\sigma$ and valley $\tau$.

\section{Characterization of the Fock space\label{sec:app_fock_space}}

\subsection{Subspace 1: Single electron in the dot, i.e. N=1}
\noindent The $N=1$ sub-matrix $H_1$ of the Hamiltonian ~\eqref{eq:fermi_hubbard_hamiltonian} is a $4\times4$ matrix as described in Tab.~\ref{tab:app:H1}.
\begin{table*}[t]
    \centering
    \caption{$H_1$, $N=1$ submatrix of the Hamiltonian}
    \begin{tabular}{|c|c c c c|}
    \hline
         & $\pmb{\ket{\kup}}$
         & $\pmb{\ket{\kpup}}$
         & $\pmb{\ket{\kdown}}$ 
         & $\pmb{\ket{\kpdown}}$\\
    \hline 
        $\pmb{\ket{\kup}}$ 
        & $\frac{1}{2}\Delta_{so}+\frac{1}{2}(g_s+g_v)\mu_B B_\perp$ 
        & $t_v$ 
        & $\frac{1}{2}g_s\mu_B B_\parallel$
        & 0 \\
        
        $\pmb{\ket{\kpup}}$ 
        & $t_v$ 
        & $-\frac{1}{2}\Delta_{so}+\frac{1}{2}(g_s-g_v)\mu_B B_\perp$ 
        & 0 
        & $\frac{1}{2}g_s\mu_B B_\parallel$ \\
        
        $\pmb{\ket{\kdown}}$ 
        & $\frac{1}{2}g_s\mu_B B_\parallel$ 
        & 0 
        & $-\frac{1}{2}\Delta_{so}+\frac{1}{2}(-g_s+g_v)\mu_B B_\perp$ 
        & $t_v$ \\
        
        $\pmb{\ket{\kpdown}}$
        & 0 
        & $\frac{1}{2}g_s\mu_B B_\parallel$ 
        & $t_v$ 
        & $\frac{1}{2}\Delta_{so}+\frac{1}{2}(-g_s-g_v)\mu_B B_\perp$ \\
    \hline
    \end{tabular}
    \label{tab:app:H1}
\end{table*} 

\; \\
\noindent \emph{Case 1: Only out-of-plane field, $(B_\parallel=0)$}\newline
\noindent From diagonalizing, we get eigenstates as

\begin{subequations}
\begin{align}
E_3^{0(1)} &= \mp\frac{1}{2}g_s \mu_{B} B_{\perp} - \sqrt{\bigg(\frac{1}{2}\Delta_{so} \mp \frac{1}{2}g_{v}\mu_{B}B_{\perp}\bigg)^2 + t_v^2}\\
E_3^{2(3)} &= \pm\frac{1}{2}g_s \mu_{B} B_{\perp} + \sqrt{\bigg(\frac{1}{2}\Delta_{so} \pm \frac{1}{2}g_{v}\mu_{B}B_{\perp}\bigg)^2 + t_v^2} 
\end{align}
\label{eq:n1perp} 
\end{subequations}

The order of the states (and the superscript label) as per the above equations changes with the value of $B_{\perp}$ at the anti crossing point given by $B_{t_1}$ in ~\eqref{eq:B_t1} and the crossing point given by $B_{t_2}$ in ~\eqref{eq:B_t2}. \\

\noindent \emph{Case 2: Only in-plane field, $(B_\perp=0)$}\newline
\noindent From diagonalizing, we get eigenstates as
\begin{subequations}
\begin{align}
E_3^{0(1)} &= -\sqrt{\frac{1}{4}\Delta_{so}^2 + \bigg(\frac{1}{2}g_s\mu_B B_{\parallel} \pm t_v\bigg)^2} \\
E_3^{2(3)} &= \sqrt{\frac{1}{4}\Delta_{so}^2 + \bigg(\frac{1}{2}g_s\mu_B B_{\parallel} \mp t_v\bigg)^2}  
\end{align}
\label{eq:n1par}
\end{subequations}

Fig.~\ref{fig:fig_2} shows the energy of these eigenstates given by ~\eqref{eq:n1perp} and ~\eqref{eq:n1par} as a function of the external magnetic field $B$ for various combinations of $\Delta_{so}$ and $t_v$. \\
\newline
Further, if we define an effective in plane $g$-factor, $g_\parallel$, for splitting of a Kramer pair in an in-plane magnetic field, given by 
\begin{equation}
    \frac{1}{2}g_\parallel \mu_B B_\parallel = \sqrt{\frac{1}{4}\Delta_{so}^2 + (\frac{1}{2}g_s\mu_B B_{\parallel} \pm t_v)^2}
\end{equation}
then solving for $g_\parallel$ for weak magnetic field $(B_\parallel\approx 0)$, gives us
\begin{equation}
    g_\parallel = \pm 2g_s\frac{t_v}{\Delta_{so}}\frac{1}{\sqrt{1+(2t_v/\Delta_{so})^2}}
\end{equation}
If $t_v\ll\Delta_{so}$, then
\begin{equation}
    g_\parallel = \pm 2g_s\frac{t_v}{\Delta_{so}}
    \label{Palyi eq}
\end{equation}
Equation~\eqref{Palyi eq} matches with the results obtained by G. Széchenyi, \textit{et al.}, (2018) \cite{szechenyi2018impurity} for the same conditions.

\subsection{Subspace 2: Two electrons in the dot, i.e. N=2}

\noindent The $N=2$ sub-matrix $H_2$ of the Hamiltonian ~\eqref{eq:fermi_hubbard_hamiltonian} is a $6\times6$ matrix as described in Tab.~\ref{tab:app:H2}. 

\begin{table*}[t]
    \centering
    \caption{$H_2$, $N=2$ submatrix of the Hamiltonian}
    \begin{tabular}{|c|@{\hskip 0.2in}c@{\hskip 0.2in}c@{\hskip 0.2in}c@{\hskip 0.2in}c@{\hskip 0.2in}c@{\hskip 0.2in}c|}
    \hline
         & $\pmb{\ket{\kup\kpup}}$ 
         & $\pmb{\ket{\kup\kdown}}$
         & $\pmb{\ket{\kup\kpdown}}$ 
         & $\pmb{\ket{\kpup\kdown}}$ 
         & $\pmb{\ket{\kpup\kpdown}}$ 
         & $\pmb{\ket{\kdown\kpdown}}$ \\  
    \hline 
        $\pmb{\ket{\kup\kpup}}$ 
        & $g_s\mu_B B_\perp$ 
        & 0 
        & $\frac{1}{2}g_s\mu_B B_\parallel$ 
        & $-\frac{1}{2}g_s\mu_B B_\parallel$ 
        & 0 
        & 0 \\

        $\pmb{\ket{\kup\kdown}}$
        & 0 
        & $g_v\mu_B B_\perp$ 
        & $t_v$ 
        & $t_v$ 
        & 0 
        & 0 \\

        $\pmb{\ket{\kup\kpdown}}$ 
        & $\frac{1}{2}g_s\mu_B B_\parallel$ 
        & $t_v$ 
        & $\Delta_{so}$ 
        & 0 
        & $t_v$ 
        & $\frac{1}{2}g_s\mu_B B_\parallel$ \\

        $\pmb{\ket{\kpup\kdown}}$ 
        & $-\frac{1}{2}g_s\mu_B B_\parallel$ 
        & $t_v$ 
        & 0 
        & $-\Delta_{so}$ 
        & $t_v$ 
        & $-\frac{1}{2}g_s\mu_B B_\parallel$ \\

        $\pmb{\ket{\kpup\kpdown}}$ 
        & 0 
        & 0 
        & $t_v$ 
        & $t_v$ 
        & $-g_v\mu_B B_\perp$ 
        & 0 \\

        $\pmb{\ket{\kdown\kpdown}}$ 
        & 0 
        & 0 
        & $\frac{1}{2}g_s\mu_B B_\parallel$ 
        & $-\frac{1}{2}g_s\mu_B B_\parallel$ 
        & 0 
        & $-g_s\mu_B B_\perp$ \\
    \hline
    \end{tabular}
    \label{tab:app:H2}
\end{table*}

\;\newline
\noindent \emph{Case 1: Only out-of-plane field, $(B_\parallel=0)$}\newline 
From diagonalizing, we get
\begin{gather}
    E_2=g_s\mu_B B_\perp \label{eq:n2perp1} \\ 
    E_2=-g_s\mu_B B_\perp \label{eq:n2perp2} \\ 
    (E_2)^4-(E_2)^2(4t_v^2+g_v^2\mu_B^2 B_\perp^2+\Delta_{so}^2) + \Delta_{so}^2g_v^2\mu_B^2 B_\perp^2=0 \label{eq:n2perp}
\end{gather}
Solving~\eqref{eq:n2perp} gives us the remaining 4 solutions
\begin{multline}
    E_2= \pm \sqrt{t_v^2+\bigg(\frac{1}{2}\Delta_{so}+\frac{1}{2}g_v\mu_B B_\perp\bigg)^2} \\
    \pm \sqrt{t_v^2+\bigg(\frac{1}{2}\Delta_{so}-\frac{1}{2}g_v\mu_B B_\perp\bigg)^2}  
    \label{eq:n2perp3}
\end{multline}
Hence, we can see that~\eqref{eq:n2perp1},~\eqref{eq:n2perp2} and~\eqref{eq:n2perp3} give the six eigenstates of the $N=2$ subspace. With some algebra, we can show that the six possible additive combinations of the four $N=1$ eigenstates given in ~\eqref{eq:n1perp} results in the six eigenstates of $N=2$ subspace given in~\eqref{eq:n2perp1},~\eqref{eq:n2perp2} and~\eqref{eq:n2perp3} as illustarted below,\\
\begin{subequations}
\begin{align}
E_2^0 &= E_1^0 + E_1^1\\
E_2^1 &= E_1^0 + E_1^2\\
E_2^2 &= E_1^0 + E_1^3\\
E_2^3 &= E_1^1 + E_1^2\\
E_2^4 &= E_1^1 + E_1^3\\
E_2^5 &= E_1^2 + E_1^3\\
\end{align}
\label{eq:n2_n1_perp} 
\end{subequations}
\noindent The exact order of the combination may be different based on the specific values of $E_i^j$\\
\;\\
\noindent \emph{Case 2: Only in-plane field,$(B_\perp=0)$}\newline
From diagonalizing, we get
\begin{gather}
    (E_2)^2=0 \label{eq:n2par1} \\ 
    (E_2)^4-(E_2)^2(4t_v^2+g_s^2\mu_B^2 B_\parallel^2+\Delta_{so}^2) + 4t_v^2g_s^2\mu_B^2 B_\parallel^2=0 \label{eq:n2par}
\end{gather}
Solving~\eqref{eq:n2par} gives us the remaining 4 solutions
\begin{multline}
    E_2=\pm \sqrt{\frac{1}{4}\Delta_{so}^2+\bigg(\frac{1}{2}g_s\mu_B B_\parallel+t_v\bigg)^2} \\
    \pm \sqrt{\frac{1}{4}\Delta_{so}^2+\bigg(\frac{1}{2}g_s\mu_B B_\parallel-t_v\bigg)^2}
    \label{eq:n2par2} 
\end{multline} 
Hence, similar to the previous case, we can see that~\eqref{eq:n2par1} and~\eqref{eq:n2par2} give the six eigenstates of the $N=2$ subspace, with two of the states being degenerate. With some algebra, we can show that the six possible additive combinations of the four $N=1$ eigenstates given in~\eqref{eq:n1par} results in the six eigenstates of $N=2$ subspace given in~\eqref{eq:n2par1} and~\eqref{eq:n2par2}, similar to ~\eqref{eq:n2_n1_perp}.

\subsection{Subspace 3: Three electrons in the dot, i.e. N=3}
The $N=3$ submatrix $H_3$ of the Hamiltonian~\eqref{eq:fermi_hubbard_hamiltonian} is a $4\times4$ matrix as described in Tab.~\ref{tab:app:H3}. It is similar to the $N=1$ submatrix with a few sign changes. 
\begin{table*}[t]
    \centering
    \caption{$H_3$, $N=3$ submatrix of the Hamiltonian}
    \begin{tabular}{|c|c c c c|}
    \hline
         & $\pmb{\ket{\kpup\kdown\kpdown}}$ 
         & $\pmb{\ket{\kup\kdown\kpdown}}$
         & $\pmb{\ket{\kup\kpup\kpdown}}$
         & $\pmb{\ket{\kup\kpup\kdown}}$ \\ 
    \hline 
        $ \pmb{\ket{\kpup\kdown\kpdown}}$
        & $-\frac{1}{2}\Delta_{so}-\frac{1}{2}(g_s+g_v)\mu_B B_\perp$ 
        & $t_v$ 
        & $-\frac{1}{2}g_s\mu_B B_\parallel$
        & 0 \\
        
        $\pmb{\ket{\kup\kdown\kpdown}}$
        & $t_v$ 
        & $\frac{1}{2}\Delta_{so}+\frac{1}{2}(-g_s+g_v)\mu_B B_\perp$
        & 0 
        & $-\frac{1}{2}g_s\mu_B B_\parallel$ \\
        
        $\pmb{\ket{\kup\kpup\kpdown}}$ 
        & $-\frac{1}{2}g_s\mu_B B_\parallel$ 
        & 0 
        & $\frac{1}{2}\Delta_{so}+\frac{1}{2}(g_s-g_v)\mu_B B_\perp$  
        & $t_v$ \\
        
        $\pmb{\ket{\kup\kpup\kdown}}$ 
        & 0 
        & $-\frac{1}{2}g_s\mu_B B_\parallel$ 
        & $t_v$ 
        & $-\frac{1}{2}\Delta_{so}+\frac{1}{2}(g_s+g_v)\mu_B B_\perp$ \\
    \hline
    \end{tabular}
    \label{tab:app:H3}
\end{table*}

\;\\
\noindent \emph{Case 1: Only out-of-plane field, $(B_\parallel=0)$}\newline
\noindent From diagonalizing, we get eigenstates as
\begin{subequations}
\begin{align}
E_3^{0(1)} &= \mp\frac{1}{2}g_s \mu_{B} B_{\perp} - \sqrt{\bigg(\frac{1}{2}\Delta_{so} \pm \frac{1}{2}g_{v}\mu_{B}B_{\perp}\bigg)^2 + t_v^2}\\
E_3^{2(3)} &= \pm\frac{1}{2}g_s \mu_{B} B_{\perp} + \sqrt{\bigg(\frac{1}{2}\Delta_{so} \mp \frac{1}{2}g_{v}\mu_{B}B_{\perp}\bigg)^2 + t_v^2} 
\end{align}
\label{eq:n3perp} 
\end{subequations}

\noindent The four solutions of the $N=3$ subspace given by ~\eqref{eq:n3perp} are the negative of the four solutions for the $N=1$ subspace given by ~\eqref{eq:n1perp}. This is expected as $H_3$ = $H_1 \times-I$, when $B_{\parallel}=0$.\\   

\noindent \emph{Case 2: Only in-plane field, $(B_\perp=0)$}\newline
\noindent From diagonalizing, we get eigenstates as
\begin{subequations}
\begin{align}
E_3^{0(1)} &= -\sqrt{\frac{1}{4}\Delta_{so}^2 + \bigg(\frac{1}{2}g_s\mu_B B_{\parallel} \pm t_v\bigg)^2} \\
E_3^{2(3)} &= \sqrt{\frac{1}{4}\Delta_{so}^2 + \bigg(\frac{1}{2}g_s\mu_B B_{\parallel} \mp t_v\bigg)^2}  
\end{align}
\label{eq:n3par}
\end{subequations}
The four solutions for the $N=3$ subspace are exactly the same as the solutions for $N=1$ eigenstates given in ~\eqref{eq:n1par} when $B_{\perp}=0$.

\section{Transport formulation\label{app_transport}} \label{sec:app2}
The total current through the quantum dot in the setup depicted in Fig.~\ref{fig:fig_1}(c)(inset) results from a complex interplay of the probability of occupation of each of the 16 eigenstates and the rates of transition between them. To tackle this problem, we extend the master equation prevalent in literature \cite{Beenakker,BM_1,BM_2,Mukherjee_2023} to our model.  We use $P_N^i$ to denote the probability of occupancy of the state $\ket{N,i}$ and $R_{(N_1,i)\rightarrow(N_2,j)}$ to denote the rate of transition from the state $\ket{N_1,i}$ to the state $\ket{N_2,j}$ by virtue of injection or removal of an electron from the source(drain). Henceforth, we shall use the index $\alpha$ for the source ($\alpha=L$) or the drain ($\alpha=R$). The probabilities $P_N^i$ evolve over time as
\begin{align}
    \dot{P}_N^i &= \sum_j \left[R_{(N\pm1,j)\rightarrow(N,i)}P_{N\pm1}^j-R_{(N,i)\rightarrow(N\pm1,j)}P_N^i\right],\label{eq:master_equation}
\end{align}
where 
\begin{align}
    R_{(N_1,i)\rightarrow(N_2,j)}=\sum_{\alpha\in\{L,R\}}R^{\alpha}_{(N_1,i)\rightarrow(N_2,j)} 
\end{align}
The rates depend on the transition matrix elements. We consider transport in the first order so that terms are non-zero if and only if $|N_1-N_2|=1$. We express the rates in terms of the matrix elements as

\begin{subequations}
\begin{alignat}{2}
    R^{\alpha}_{(N,i)\rightarrow(N-1,j)}&=\Gamma_{Nr}^{{\alpha} ij}\left[1-f\left(\frac{\epsilon_{N}^{ij}-\mu_{\alpha}}{k_BT}\right)\right]\\
    R^{\alpha}_{(N,i)\rightarrow(N+1,j)}&=\Gamma_{Na}^{{\alpha} ij}f\left(\frac{\epsilon_{N+1}^{ji}-\mu_{\alpha}}{k_BT}\right),
\end{alignat}
\end{subequations}
where $k_B$ is the Boltzmann's constant, $T$ is the temperature of the system (we assume this to be uniform across the entire system), $f$ is the Fermi-Dirac distribution, and $\Gamma_{Nr(a)}^{{\alpha} ij}$ is the matrix element for the removal(addition) of an electron defined as
\begin{subequations}
\begin{alignat}{2}
    \Gamma_{Nr}^{{\alpha} ij}&=\sum_{\sigma\tau}\gamma_{\alpha}\left|\bra{N,i}\hat{c}_{\sigma\tau}\ket{N+1,j}\right|^2,\\
    \Gamma_{Na}^{{\alpha} ij}&=\sum_{\sigma\tau}\gamma_{\alpha}\left|\bra{N,i}\hat{c}_{\sigma\tau}^\dagger\ket{N-1,j}\right|^2.
\end{alignat}
\end{subequations}

The coefficient $\gamma_{L(R)}$, illustrated in Fig.~\ref{fig:fig_1}(c)(inset), represents the contact coupling rates of the SQD with the source(drain) \cite{meir1991transport}. 

To obtain the total current, one has to solve \eqref{eq:master_equation} for $\dot{P}_N^i=0, \forall N,i$ under the constraint $\sum_{N,i}P_N^i=1$. The expression for current is given by \cite{PhysRevB.76.035432}
\begin{align}
    I = \frac{e}{h}\sum_{N=1}^4\sum_{\langle i,j\rangle}\left[R^L_{(N-1,j)\rightarrow(N,i)}P_{N-1}^j-R^L_{(N,i)\rightarrow(N-1,j)}P_N^i\right].\label{eq:current}
\end{align}

\bibliography{references}

\end{document}